\documentclass[12pt]{article}
\usepackage[latin1]{inputenc}

\usepackage{amsmath}
\usepackage{amsfonts}
\usepackage{amssymb}
\usepackage{graphicx}
\usepackage{geometry}
\usepackage{amssymb,epsfig,subfigure}
\usepackage{hyperref}

\pdfoutput=1

\makeatletter
\renewcommand\section{\@startsection {section}{1}{\z@}%
                                 {-3.5ex \@plus -1ex \@minus -.2ex}
                                   {2.3ex \@plus.2ex}%
                                   {\normalfont\large\bfseries}}
\renewcommand\subsection{\@startsection{subsection}{2}{\z@}%
                                   {-3.25ex\@plus -1ex \@minus -.2ex}%
                                     {1.5ex \@plus .2ex}%
                                     {\normalfont\bfseries}}
\renewcommand\subsubsection{\@startsection{subsubsection}{3}{\z@}%
                                   {-3.25ex\@plus -1ex \@minus -.2ex}%
                                     {1.5ex \@plus .2ex}%
                                     {\normalfont\itshape}}
\makeatother

\def\pplogo{\vbox{\kern-\headheight\kern -29pt
\halign{##&##\hfil\cr&{\ppnumber}\cr\rule{0pt}{2.5ex}&\ppdate\cr}}}
\makeatletter
\def\ps@firstpage{\ps@empty \def\@oddhead{\hss\pplogo}%
  \let\@evenhead\@oddhead 
}
\def\maketitle{\par
 \begingroup
 \def\thefootnote{\fnsymbol{footnote}}
 \def\@makefnmark{\hbox{$^{\@thefnmark}$\hss}}
 \if@twocolumn
 \twocolumn[\@maketitle]
 \else \newpage
 \global\@topnum\z@ \@maketitle \fi\thispagestyle{firstpage}\@thanks
 \endgroup
 \setcounter{footnote}{0}
 \let\maketitle\relax
 \let\@maketitle\relax
 \gdef\@thanks{}\gdef\@author{}\gdef\@title{}\let\thanks\relax}
\makeatother

\numberwithin{equation}{section}

\renewcommand{\dag}{\dagger}

\newcommand{\be}{\begin{equation}}
\newcommand{\bea}{\begin{eqnarray}}
\newcommand{\ee}{\end{equation}}
\newcommand{\eea}{\end{eqnarray}}

\newcommand{\mc}{\mathcal}

\newcommand{\tr}{{\rm tr}}
\renewcommand{\t}{\tilde}
\newcommand{\muphi}{\mu_\phi}

\begin{document}

\setcounter{page}0
\def\ppnumber{\vbox{\baselineskip14pt
}}
\def\ppdate{\footnotesize{SLAC-PUB-14430}} \date{}

\author{Anson Hook, Gonzalo Torroba\\
[7mm]
{\normalsize  \it SLAC and Department of Physics, }\\
{\normalsize \it Stanford University, Stanford, CA 94309, USA}\\
[3mm]}

\bigskip
\title{\bf  A microscopic theory of gauge mediation
\vskip 0.5cm}
\maketitle

\begin{abstract}
We construct models of indirect gauge mediation where the dynamics responsible for breaking supersymmetry simultaneously generates a weakly coupled subsector of messengers. This provides a microscopic realization of messenger gauge mediation where the messenger and hidden sector fields are unified into a single sector. The UV theory is SQCD with massless and massive quarks plus singlets, and at low energies it flows to a weakly coupled quiver gauge theory. One node provides the primary source of supersymmetry breaking, which is then transmitted to the node giving rise to the messenger fields. These models break R-symmetry spontaneously, produce realistic gaugino and sfermion masses, and give a heavy gravitino.
\end{abstract}
\bigskip
\newpage

\tableofcontents

\vskip 1cm

\section{Introduction}\label{sec:intro}

In gauge-mediated supersymmetry breaking, supersymmetry is broken in a hidden sector and then communicated to the supersymmetric Standard Model (SSM) via gauge interactions~\cite{Dine:1993yw,Dine:1994vc,Dine:1995ag}. The gauge mediation mechanism may be `direct' or `indirect' depending on the existence of an additional messenger sector that does not participate in supersymmetry breaking.

In direct mediation~\cite{Affleck:1984xz} the hidden sector has a flavor symmetry, part of which is identified with the SM gauge group $G_{SM}$. The visible soft parameters are generated by loops of fields which we may call `messengers,' but the important property of direct mediation is that these messengers are an integral part of the supersymmetry breaking mechanism. If the couplings to the messengers are set to zero, supersymmetry is restored~\cite{Dine:2007dz}. The recent discovery of metastable vacua in supersymmetric QCD (SQCD) by Intriligator, Seiberg and Shih (ISS)~\cite{Intriligator:2006dd} has greatly simplified the construction of models of direct mediation.\footnote{The literature in this direction is already vast; we refer the reader to the nice reviews~\cite{Intriligator:2007cp,Kitano:2010fa,Dine:2010cv}.} 

In models of indirect mediation the messenger sector can be decoupled without affecting supersymmetry breaking. The prototype for indirect mediation is minimal gauge mediation (MGM); see~\cite{Giudice:1998bp} for a detailed review and references.  In these models, the messenger sector contains fields transforming under a vector-like representation of $G_{SM}$ that couple weakly to a hidden sector spurion $X$ with nonzero expectation value and F-term, $W=X \Phi \t \Phi$. More general interactions may also be added (see e.g.~\cite{Dumitrescu:2010ha}), and we will refer to this mechanism as messenger gauge mediation.

Both frameworks have their own advantages. Models of messenger gauge mediation insulate the hidden sector from the SSM, which is very useful for achieving perturbative gauge coupling unification and gives more flexibility in the generation of soft parameters. However, the existence of intermediate subsectors to communicate supersymmetry and the ad-hoc nature of messengers and their interactions are not very appealing. Theories of direct mediation are in this sense simpler, avoiding an intermediate messenger sector. However, this requirement tightly constrains the hidden sector and tends to produce Landau poles because the flavor symmetry has to be large enough to contain $G_{SM}$.

It would be useful to find models of gauge mediation that combine the attractive features of direct and indirect mediation. This requires a microscopic realization of indirect mediation. Unfortunately, it appears to be quite hard to build a complete and consistent model of messenger gauge mediation. Many of the results so far are at an effective level, treating the hidden sector as a spurion.\footnote{A general parametrization was given in~\cite{Meade:2008wd}.} A UV completion of gauge mediation is also important for addressing $\mu/B_\mu$. Our work will improve this situation by providing a concrete microscopic realization of gauge mediation with messengers. We will see that the dynamics of SQCD and Seiberg duality~\cite{Seiberg:1994pq} are central for achieving these goals, much as they were for direct mediation.

If the supersymmetry breaking mechanism and the messenger fields have a common UV origin, both mediation schemes could be merged. The unification of hidden and messenger sectors is also motivated by duality: a model of dynamical supersymmetry breaking with a messenger sector and a model of direct mediation could be `magnetic' and `electric'  dual descriptions of a single microscopic gauge theory.  Models constructed in this paper will have this property.

The goal of this work is to construct models where the gauge theory responsible for breaking supersymmetry, simultaneously generates weakly coupled messengers. The main consequence of this is the unification of the hidden and messenger sectors. As in direct mediation, no messenger fields have to be added by hand to the theory; but now the messengers do not play an important role in the breaking of supersymmetry despite their direct couplings to the hidden fields.
Our approach is based on SQCD in the free magnetic phase, and uses the ISS mechanism to break supersymmetry~\cite{Intriligator:2006dd}. This will give a concrete realization of a duality between direct and indirect mediation. From the point of view of the electric theory our construction will amount to a model of direct mediation, while the magnetic description will yield a model with a messenger subsector weakly interacting with an O'Raifeartaigh model.

Having a UV completion of messenger gauge mediation will allow us to go beyond the effective description of messenger gauge mediation. While from a bottom-up viewpoint these couplings could appear to be arbitrary, embedding them into a consistent microscopic theory will introduce various constraints and will teach us lessons about the way in which the messenger fields may interact. Next, we will use these tools to construct models of gauge mediation that achieve gauge coupling unification and produce a fully realistic soft spectrum for the SSM. The second part of the paper will be devoted to this and will present the basic phenomenology of our constructions.

\subsection{Basic mechanism and overview}\label{subsec:intro-sqcd}

Before proceeding to a detailed analysis, let us explain our basic strategy in a simple setup. Consider $SU(N_c)$ SQCD with $N_f=N_c+1$ flavors $(Q, \t Q)$. Below the dynamical scale $\Lambda$, the theory admits a magnetic description in terms of mesons and baryons with superpotential~\cite{Seiberg:1994pq}
\be\label{eq:Wmag1}
W= \frac{1}{\Lambda^{2N_c-1}}\left(\det M+\tr\, B M \t B \right)\,.
\ee
This equation suggests building a model of gauge mediation by identifying the supersymmetry breaking spurion with the meson $M$, and the baryons with the messengers. Supersymmetry can be broken as in the ISS model by adding a common mass to the electric quarks, $W_{el}= m Q \t Q$. For $|m| \ll |\Lambda|$, the magnetic theory becomes
\be
W= m \tr\,M + \frac{1}{\Lambda^{2N_c-1}}\tr\, B M \t B+\frac{1}{\Lambda^{2N_c-1}}\det M\,.
\ee
Near the origin of field space, the last term may be ignored, and the first two terms break supersymmetry by the rank condition~\cite{Intriligator:2006dd}.

A model of gauge mediation is obtained by weakly gauging a subgroup $G_{SM} \subset SU(N_f)$ and identifying it with the SM gauge group. As it stands, this is a model of direct mediation, with the messengers $(B, \t B)$ participating directly in supersymmetry breaking. For our purposes, we want to obtain a model of indirect mediation while preserving the nice fact that the messengers are `dynamically' generated. The solution to the problem is to allow for different electric quark masses. As we explain below, the scale of supersymmetry breaking is determined by the rank condition, where the maximal number of (largest) linear terms are canceled by expectation values of $(B, \t B)$. The messengers are the baryons with the lighter masses.

More concretely, let us assume that there are $N_{f,1}$ massive and $N_{f,0}$ massless quarks.\footnote{SQCD with massless and massive quarks has a rich history of applications. Its potential was realized by~\cite{Franco:2006es}, in connection with branes at singularities. A two-loop field theory analysis, which will be used in this work, was performed in~\cite{Giveon:2008wp}. The theory at two-loops has a runaway that will be stabilized in \S \ref{subsec:2-loop}. Some recent applications to supersymmetry breaking, R-symmetry breaking and $\mu/B_\mu$ were considered in~\cite{Giveon:2008ne,SchaferNameki:2010mg}.} The meson matrix can be split accordingly into $(M_{11},M_{10},M_{01}, M_{00})$, and similarly for the baryons, $B=(B_1, B_0)$.
The magnetic theory then gives
\begin{itemize}
\item[i)] a sector $(M_{11}, B_1, \t B_1)$ that breaks supersymmetry spontaneously for $N_{f,1}>1$;
\item[ii)] a sector $(M_{00}, B_0, \t B_0)$ that does not participate in supersymmetry breaking;
\item[iii)] fields $(M_{10}, M_{01})$ that couple both sectors.
\end{itemize}
By weakly gauging $G_{SM}\subset SU(N_{f,0})$, $(B_0, \t B_0)$ can play the role of messenger fields if adequate supersymmetry breaking properties can be induced on $M_{00}$. This will be accomplished by adding a small superpotential deformation for $M_{00}$, and balancing it against two-loop corrections.

Summarizing, SQCD with massless and massive flavors and adequate superpotential deformations can give rise, in the IR, to a supersymmetry breaking hidden sector and a weakly coupled messenger sector. The messenger fields are composites of the same confining dynamics that breaks supersymmetry. In this work we will study this theory in detail and construct realistic models of messenger gauge mediation.

\vskip 2mm

The paper is organized as follows. First, in \S \ref{sec:different} we study SQCD with massless and massive quarks focusing on the generation of weakly coupled subsectors. The theory flows in the IR to a 3-node quiver with rich dynamics that includes supersymmetry breaking, one loop stabilizing effects and two loop destabilizing effects. We find two interesting regimes, corresponding to $M_{00}$ (introduced above) being stabilized at small or large values. While this approach is general, in \S \ref{sec:dyn-mess} we focus on the construction of theories of gauge mediation with `dynamical' messengers. Implementing dynamical messengers requires supplementing the electric theory with spectator fields in order to decouple certain mesons that would otherwise induce tachyonic sfermions. We study the supersymmetry breaking vacuum and the properties of the messenger sector.  Implications for an effective theory of messenger gauge mediation are also presented. 
The phenomenology of the models is considered in \S \ref{sec:pheno}. We give parameter ranges leading to TeV scale gauginos and sfermions and find in general a heavy gravitino that with nonstandard cosmology can account for the dark matter density. \S \ref{sec:concl} contains our conclusions and some future directions.

\section{Weakly coupled subsectors from SQCD}\label{sec:different}

We first discuss how weakly coupled subsectors arise from SQCD with different quark masses. This analysis will be general since it could have different applications. Next, in \S \ref{sec:dyn-mess} we will focus on models of gauge mediation with messengers. The results of this section have already appeared before (see e.g.~\cite{Franco:2006es,Giveon:2008wp,Giveon:2008ne}), but our motivation and approach are, as far as we know, new.

\subsection{Electric and magnetic descriptions}\label{subsec:e-m}

Consider $SU(N_c)$ SQCD with $N_f$ flavors, in the free magnetic range $N_c+1 \le N_f < \frac{3}{2}N_c$. $N_{f,0}$ of the electric quarks are massless, while
$$
N_{f,1}=N_f-N_{f,0}
$$ 
quarks have mass $m$,
\be
W_{el}= m \sum_{i=1}^{N_{f,1}} Q_i \t Q_i\,.
\ee
The masses are taken to be much smaller than the dynamical scale, $|m| \ll |\Lambda|$. As we shall explain soon, the number of massless quarks cannot exceed $N_c-1$ in order to break supersymmetry. The mass terms break the anomaly-free nonabelian flavor group $SU(N_f)$ down to $SU(N_{f,1}) \times SU(N_{f,0})$. 

The Seiberg-dual magnetic theory~\cite{Seiberg:1994pq} has gauge group $SU(\t N_c \equiv N_f -N_c)$, $N_f$ magnetic quarks $(q, \t q)$ and $N_f^2$ singlets corresponding to the (normalized) mesons of the electric theory,
$$
\Phi_{ij}= \frac{(Q_i \t Q_j)}{\Lambda}\,.
$$
The superpotential is
\be
W_{mag}=\tr(\Lambda m\,\Phi)+ h \tr(q \Phi \t q)\,.
\ee
The matter content with the anomaly-free nonabelian symmetries is
\begin{center}
\begin{tabular}{c|cccc}
&$SU(\t N_c)_G$&$SU(N_{f,1})$&$SU(N_{f,0})$  \\
\hline
&&&\\[-12pt]
$q_1$&$\Box$&$\overline \Box$&$1$  \\
$\t q_1$&$\overline \Box$&$ \Box$&$1$  \\
$q_0$&$\Box$&$1$&$\overline \Box$  \\
$\t q_0$&$\overline\Box$&$1$&$\Box$  \\
$\Phi_{11}$&$1$&$\textrm{adj}+1$&$1$  \\
$\Phi_{00}$&$1$&$1$&$\textrm{adj}+1$  \\
$\Phi_{10}$&$1$&$\Box$&$\overline \Box$  \\
$\Phi_{01}$&$1$&$\overline\Box$&$ \Box$  
\end{tabular}
\end{center}
Abelian symmetries will be discussed shortly. In the free magnetic range the magnetic theory has a Landau pole at a scale $\t \Lambda$, which is here identified with $\Lambda$. Above this scale, the theory is UV-completed by the electric description.

\begin{figure}[h!]
\centering
\includegraphics[width=4in]{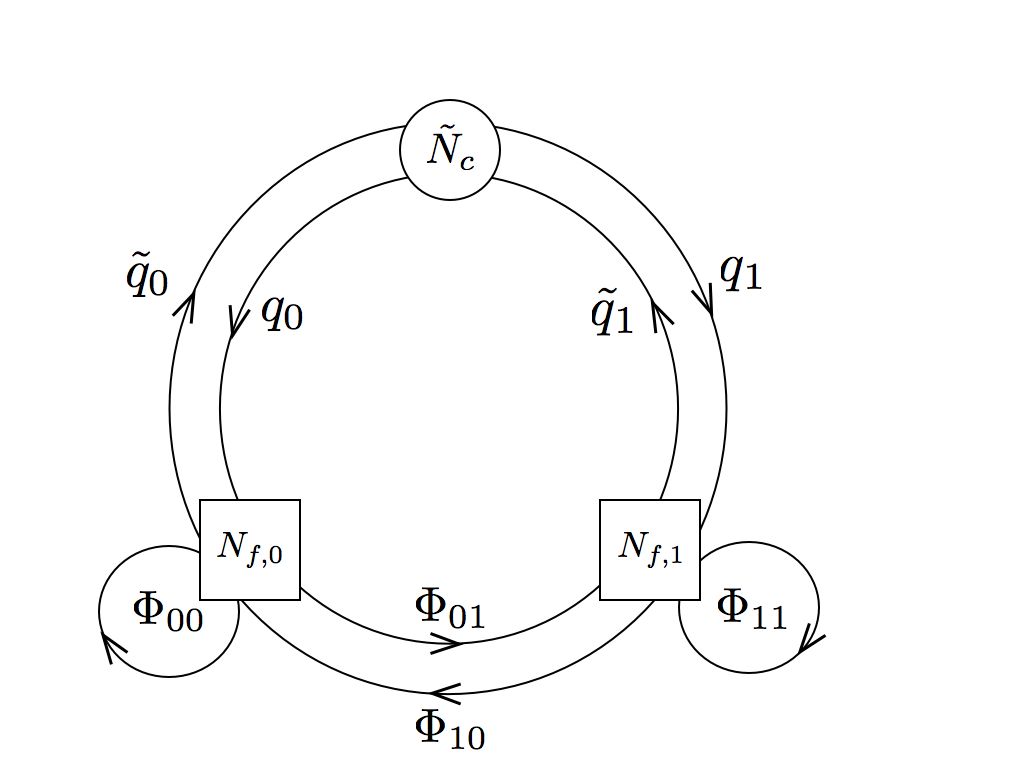}
\caption{\small{Quiver gauge theory from SQCD with massless and massive quarks. An arrow coming out from a node denotes a field transforming in the fundamental representation. Circular nodes are gauge symmetries while rectangles represent flavor symmetries.}}
\label{fig:quiver1}
\end{figure}

Therefore, having both massive and massless nodes leads, in the IR, to a quiver gauge theory with three nodes corresponding to $SU(\t N_c)_G \times SU(N_{f,0})\times SU(N_{f,1})$. The matter content can be usefully represented by the quiver diagram of Figure~\ref{fig:quiver1}. In terms of this decomposition, the magnetic superpotential neglecting nonperturbative corrections is
\be\label{eq:Wgen}
W_{mag}=\left[ - h \mu^2 \,\tr \Phi_{11} + h \tr(q_1 \Phi_{11} \t q_1) \right]+ h \tr(q_1 \Phi_{10} \t q_0 + q_0 \Phi_{01} \t q_1)+ h \tr (q_0 \Phi_{00} \t q_0)\,,
\ee
where $-h \mu^2 \approx m \Lambda$. The superpotential parameters will be chosen to be real. The K\"ahler potential is approximately canonical in the fields $\Phi$, $q$ and $\t q$.

The theory with superpotential (\ref{eq:Wgen}) naturally splits into a supersymmetry breaking $SU(N_{f,1})$ node (the terms inside the square brackets in (\ref{eq:Wgen})), a second tree level supersymmetric $SU(N_{f,0})$ node, and interaction terms containing bifundamentals under these flavor symmetry groups. While in this work we focus on the simplest case of massless and massive flavors, we point out that the generalization
to different quark masses produces a quiver with a central node corresponding to the magnetic gauge group and various nodes arising from the flavor symmetries left unbroken by the matrix $-h\mu^2=m \Lambda$. Each of these flavor nodes is connected to the central node via magnetic quarks and is connected to all the other flavor nodes by bifundamental meson fields and interaction terms determined from the coupling $W= h \tr(q\Phi \t q)$. The flavor node with largest $\mu$ provides the dominant source of supersymmetry breaking.  The breaking of supersymmetry and its transmission to the other nodes will be studied in the next subsection.

In this way, SQCD with different electric quark masses leads to a rich class of quivers in the IR, providing new mechanisms of supersymmetry breaking and different schemes for mediating this to the visible sector. For instance, now the SM gauge group can be embedded in different flavor groups. In this work the SSM will be connected to the hidden sector via the $N_{f,0}$ node.\footnote{The other alternatives may be interesting for other purposes and some of them have been already explored. For instance, \cite{SchaferNameki:2010mg} considered a model of direct mediation by weakly gauging $SU(N_{f,1})$ while using the $N_{f,0}$ node to solve the $\mu/B_\mu$ problem. It could be interesting to explore applications of our quiver theory to gauging $G_{SM} \subset SU(\t N_c)$, after the breaking to the diagonal described below. The analogous case but without massless quarks was studied by~\cite{Kitano:2006xg}.}

\subsection{Dynamical supersymmetry breaking}\label{subsec:dsb}

The quiver theory of Figure \ref{fig:quiver1} has very interesting dynamics, with different effects taking place at tree level, one loop and two loops. These give a weakly coupled description of highly nontrivial nonperturbative physics in the microscopic electric theory, related to condensation of monopoles. Let us understand each of these levels in turn.

First, at tree level the sector $(\Phi_{11}, q_1, \t q_1)$ breaks supersymmetry by the rank condition as long as $N_{f,1}\ge \t N_c+1$ or, equivalently, $N_{f,0}\le N_c-1$. The fields are parametrized by
\begin{equation}\label{eq:param2}
\Phi_{11}= \left(\begin{matrix} Y_{\tilde N_c \times \tilde N_c} & Z^T_{\tilde N_c \times (N_{f,1} - \tilde N_c) } \\ \tilde Z_{(N_{f,1}-\tilde N_c)  \times \tilde N_c} &X_{(N_{f,1}-\tilde N_c)  \times (N_{f,1}-\tilde N_c) }\end{matrix} \right)\,,\;\;
q^T_1=\left( \begin{matrix} \chi_{\tilde N_c \times \tilde N_c} \\ \rho_{(N_{f,1}-\tilde N_c ) \times \tilde N_c} \end{matrix}\right)\,,\;\tilde q_1=\left( \begin{matrix} \tilde \chi_{\tilde N_c \times \tilde N_c} \\ \tilde \rho_{(N_{f,1}-\tilde N_c ) \times \tilde N_c} \end{matrix}\right)\,.
\end{equation}
Recall that $N_{f,1}$ runs over the massive flavors. The bifundamentals $\Phi_{10}$ and $\Phi_{01}$ are decomposed accordingly into
\be
\Phi_{10}=\left( \begin{matrix}(Z_0)_{\tilde N_c \times N_{f,0}} \\ L_{(N_{f,1}-\tilde N_c ) \times N_{f,0}} \end{matrix}\right)\;,\;\;\Phi_{01}^T=\left( \begin{matrix}(\t Z_0)_{\tilde N_c \times N_{f,0}} \\ \t L_{(N_{f,1}-\tilde N_c ) \times N_{f,0}} \end{matrix}\right)\,.
\ee

The vacuum is characterized by
\be\label{eq:ISSvac}
\langle \chi \t \chi \rangle = \mu^2 \mathbf 1_{\t N_c \times \t N_c}\;\;,\;\;W_X = - h \mu^2\,\mathbf 1_{(N_{f,1}-\tilde N_c)  \times (N_{f,1}-\tilde N_c)}\,.
\ee
The fields $X, L, \t L, \Phi_{00}$ and  $\rm{Re}\,\tr(\chi - \t \chi))$ are massless at tree level (`pseudo-moduli'); all the other fields, except for some exactly massless Nambu-Goldstone bosons, are massive and stabilized at the origin. Notice that, unlike $X$, $\Phi_{00}$ does not have a linear term and so its F-term vanishes. This will be important for the phenomenological applications in \S\S \ref{sec:dyn-mess} and \ref{sec:pheno}.

This theory has four independent $U(1)$ symmetries~\cite{Giveon:2008wp}, and two of them will play an important role here. The first one is the $U(1)_V$ baryon number that assigns charge $+1$ to $q$ and $-1$ to $\t q$. The other important abelian symmetry is an R-symmetry, discussed in \S \ref{subsec:2-loop}.
At the origin of pseudo-moduli space the pattern of symmetry breaking is
\be\label{eq:breaking}
SU(\t N_c)_G \times SU(N_{f,1}) \times SU(N_{f,0}) \times U(1)_V \to SU(\t N_c)_D \times SU(N_{f,1}-\t N_c) \times SU(N_{f,0}) \times U(1)'\,.
\ee

\begin{figure}[htb]
\centering
\includegraphics[width=3.6in]{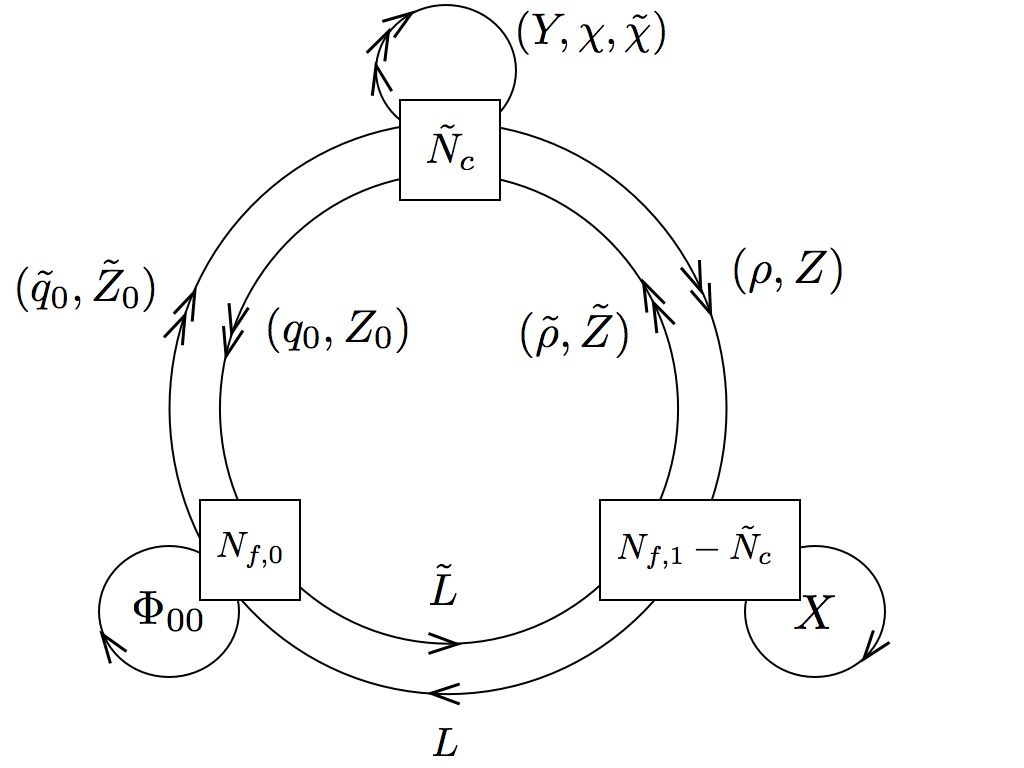}
\caption{\small{Quiver gauge theory for magnetic SQCD after supersymmetry breaking triggered by the node $SU(N_{f,1})$.}}
\label{fig:quiver2}
\end{figure}

The low energy theory after supersymmetry breaking is given by the quiver of Figure \ref{fig:quiver2}.

\subsubsection{Tree-level and one-loop spectrum}

Having understood the vacuum structure, let us analyze the tree-level spectrum of the theory. The spectrum depends on the position along pseudo-moduli space, and we start by setting all the pseudo-moduli to zero; see~\cite{Intriligator:2006dd,Essig:2008kz,Giveon:2008wp}.
\vskip 1mm
\noindent (1) \textit{The $(Y, \chi)$ sector}: $Y$ and $\chi+\t \chi$ form $\t N_c^2$ supersymmetric massive Dirac fermions with masses of order $h \mu$. The traceless part of $\chi - \t \chi$ is eaten by the super Higgs mechanism. $\rm{Re}\,\tr(\chi - \t \chi)$ is a pseudo-modulus, while $\rm{Im}\,\tr(\chi - \t \chi)$ is a NG boson associated to the breaking of $U(1)_V$. We weakly gauge $U(1)_V$ to avoid having this massless scalar. This sector is supersymmetric at tree level and will not play an important role in the following discussion.
\vskip 1mm
\noindent (2) \textit{The $(\rho, Z)$ sector}: there are $2(N_{f,1}- \t N_c) \t N_c$ Dirac fermions with masses of order $h \mu$. There are $(N_{f,1}- \t N_c)\t N_c$ complex NG bosons $\rm{Re}(\rho+ \t \rho)$ and $\rm{Im}(\rho+ \t \rho)$. These massless scalars would be phenomenologically forbidden and can be lifted either by weakly gauging part of the $SU(N_{f,1})$ symmetry or by turning on slightly different $\mu^2$'s.

The important fields for model-building come from the remaining $3(N_{f,1}- \t N_c) \t N_c$ complex scalars. These have supersymmetric masses and F-term splittings of order $h \mu$ from their direct coupling to the nonzero F-term $W_X$. They will generate the leading soft corrections for the other light fields.
\vskip 1mm
\noindent (3) \textit{The $(q_0, Z_0)$ sector}: at tree-level this is a supersymmetric sector containing $2N_{f,0} \t N_c$ superfields with masses of order $h \mu$. The fields $q_0$ will later on play the role of messengers.
\vskip 1mm
\noindent (4) The $(N_{f,1}-\t N_c)^2$ scalars $X$ and their fermionic partners are flat at tree level, with the trace $\psi_{\tr X}$ being the Goldstino.
\vskip 1mm
\noindent (5) The $(L, \t L)$ fields give $2(N_{f,1}- \t N_c)N_{f,0}$ massless supermultiplets at tree-level.
\vskip 1mm
\noindent (6) Finally, $\Phi_{00}$ also gives $N_{f,0}^2$ tree level massless supermultiplets. This field will be important for inducing F-term splittings to the messengers, via the superpotential interaction $W \supset h \tr(q_0 \Phi_{00} \t q_0)$.

The one-loop potential for the pseudo-moduli is calculated by taking them as background fields and using the Coleman-Weinberg formula~\cite{Coleman:1973jx}. From
\be
W \supset  h \mu \,\tr(\rho \t Z + Z \t \rho+ q_0 \t Z_0 + Z_0 \t q_0)+ h\, \tr \left( \rho \langle X \rangle \t \rho+ q_0 \langle \Phi_{00} \rangle \t q_0 + \rho \langle L \rangle \t q_0 + q_0 \langle \t L \rangle \t \rho\right)
\ee
one computes the fermionic and bosonic mass matrices (the later include off-diagonal terms induced by $W_X=-h \mu^2$) and integrating out these fields generates a one loop contribution
\be\label{eq:cw1}
V_{CW}^{(1)}=\frac{1}{64\pi^2} \,{\rm Str} \,M^4 \log \frac{M^2}{\Lambda_{{\rm cutoff}}^2}\,.
\ee
The finite corrections are of course independent of the cutoff $\Lambda_{\rm{cutoff}}$, which for simplicity can be chosen at the supersymmetry breaking scale $h \mu$.

An important consequence of (\ref{eq:cw1}) is a positive squared mass for $X$,
\be
m_{CW}^2 = \frac{h^2}{8 \pi^2} (\log 4-1) \t N_c (h \mu)^2
\ee
and a similar mass for the pseudo-modulus $\rm{Re}\,\tr(\chi - \t \chi)$. In what follows we suppress order one numerical factors like $\log 4 -1$.
On the other hand, $\Phi_{00}$ does not receive a one loop mass because at this order it does not couple to fields with nonsupersymmetric splittings.
The bifundamentals $(L, \t L)$ do obtain a one loop mass, which depends on the expectation value of $\Phi_{00}$. Near the origin their mass is simply given by $m_{CW}^2$, while for large $\Phi_{00}$ there is a suppression,
\be\label{eq:cwL2}
m_{L}^2\approx  \left(\frac{h \mu}{\Phi_{00}} \right)^2 m_{CW}^2\,.
\ee
The fermions also acquire nonzero masses of order the CW scale. See~\cite{Giveon:2008ne} and \S \ref{subsec:2-loop}.

\subsection{Two-loop analysis and stabilization mechanisms}\label{subsec:2-loop}

We see that the $N_{f,1}$ node provides the source of supersymmetry breaking, and this is then transmitted to the $N_{f,0}$ node starting at two loops. These turn the origin of $\Phi_{00}$ tachyonic,\footnote{For small $\Phi_{00}$ this analysis was performed in~\cite{Giveon:2008wp} using the results of~\cite{Martin:2001vx}. Results for large $\Phi_{00}$ were obtained analytically by~\cite{Giveon:2008ne} based on the wavefunction renormalization found in~\cite{Intriligator:2008fe}, and numerically using the 2-loop potential in~\cite{SchaferNameki:2010mg}. Nonperturbative effects were studied in~\cite{Franco:2006es}. } 
\be\label{eq:V2origin}
V^{(2)}_{CW}\approx- \t N_c (N_{f,1}-\t N_c)\left(\frac{h^2}{16 \pi^2} \right)^2 (h \mu)^2 \,\tr(\Phi_{00}^\dag \Phi_{00})\,,
\ee
where we are suppressing order one numerical factors. We will shortly describe how this direction is stabilized and leads to visible soft terms, but it is important to stress that the primary source of supersymmetry breaking is the $N_{f,1}$ node. In the limit $\mu \to 0$, supersymmetry is restored in the full quiver theory.

The system is analytically tractable in two regimes: at small field values when the CW potential is well-approximated by (\ref{eq:V2origin}), and in the regime
$\Phi_{00} \gg \mu$, for which the potential becomes logarithmic,
\be\label{eq:V2log}
V^{(2)}_{CW}\approx - \t N_c (N_{f,1}- \t N_c) \left(\frac{h^2}{16 \pi^2} \right)^2\,h^2 \mu^4\,\tr \log^2 \frac{\Phi_{00}^\dag \Phi_{00}}{\mu^2}\,.
\ee
We will consider these two limits, for which simple stabilization mechanisms exist. The regime of intermediate values requires a numerical analysis.

These effects are central to our goal of constructing theories of messenger gauge mediation. By adding
small superpotential deformations we will stabilize $\Phi_{00}$ at small and large values and induce nonzero F-terms that will eventually be responsible for the MSSM soft masses. As a first step, let's explain how $\Phi_{00}$ is stabilized.\footnote{These deformations also appeared in~\cite{SchaferNameki:2010mg} in the context of the $\mu/B_\mu$ problem.}

\subsubsection{Regime $\Phi_{00} \ll \mu$}

It is possible to stabilize $\Phi_{00}$ at small values if we deform the magnetic superpotential by a cubic interaction,
\be\label{eq:def1}
\Delta W_{mag}= \frac{1}{3} \lambda\,\Phi_{00}^3\,.
\ee
This operator arises from a dimension 6 perturbation $(Q \t Q)^3$ in the electric theory . An important property of the low energy theory with superpotential (\ref{eq:Wgen}) plus (\ref{eq:def1}) is that it preserves a classical R-symmetry with charges
\be
R_{\Phi_{11}}=2\;\;,\;\;R_{q_1}=R_{\t q_1}=0\;\;,\;\;R_{\Phi_{00}}=R_{q_0}=R_{\t q_0}=\frac{2}{3}\;\;,\;\;R_{\Phi_{10}}=R_{\Phi_{01}}=\frac{4}{3}\,.
\ee

Balancing (\ref{eq:def1}) against the two-loop potential (\ref{eq:V2origin}) gives
\bea
\langle \Phi_{00}^\dag \Phi_{00}\rangle &\approx& \frac{1}{2}\t N_c (N_{f,1}-\t N_c)\left(\frac{h^2}{16 \pi^2} \right)^2 \frac{h^2 \mu^2}{\lambda^2}\,\mathbf 1_{N_{f,0}\times N_{f,0}}\nonumber\\
\langle W_{\Phi_{00}} \rangle&=&  \lambda  \langle \Phi_{00}^2 \rangle \approx \frac{1}{2}\t N_c (N_{f,1}-\t N_c)\left(\frac{h^2}{16 \pi^2} \right)^2 \frac{h^2 \mu^2}{\lambda}e^{2i\rm{arg}\,(\Phi_{00})}\,\mathbf 1_{N_{f,0}\times N_{f,0}}\,.
\eea
The R-symmetry is \textit{spontaneously} broken by the expectation value of $\Phi_{00}$ and the massless phase $\rm{arg}(\Phi_{00})$ is the R-axion.\footnote{Our results on spontaneous breaking of R-symmetry at two-loops is in agreement with the analysis of~\cite{Amariti:2008uz}} We postpone the discussion of the stabilization of the axion to \S  \ref{sec:pheno}. Self-consistency of the approximation $\Phi_{00} \ll \mu$ sets an upper bound
\be\label{eq:cond1}
\frac{\lambda}{h} \gtrsim \left( \frac{1}{2}\t N_c (N_{f,1}-\t N_c)\right)^{1/2} \frac{h^2}{16 \pi^2}\,.
\ee
This means that the the nonrenormalizable scale controlling the dimension 6 operator in the electric theory cannot exceed the dynamical scale by more than a couple of orders of magnitude, depending on the size of $h$.

After $\Phi_{00}$ acquires expectation values for its lowest component and F-term, the coupling $W\supset h q_0 \Phi_{00} \t q_0$ induces supersymmetric masses and splittings for the $q_0$ fields,
\be
M \sim h |\Phi_{00}|\;\;,\;\;F \sim h \lambda |\Phi_{00}|^2\,.
\ee
Here we have ignored the other fields charged under the standard model which, in our realistic models, will be lifted by the addition of singlets.

\subsubsection{Regime $\Phi_{00} \gg \mu$}

In the limit of large field values we consider instead a quadratic deformation
\be\label{eq:def2}
\Delta W= \frac{h^2}{2}\muphi\,\Phi_{00}^2
\ee
that arises from a dimension 4 operator $(Q \t Q)^2$ in the electric theory\footnote{A similar deformation for $X$, that breaks the R-symmetry both explicitly and spontaneously, was studied in detail in~\cite{Essig:2008kz}. Note however that the models presented in our work do not break the R-symmetry explicitly, but only spontaneously. This was emphasized by~\cite{Giveon:2008ne}.}. This naturally gives $\muphi \ll \mu$ since $\mu$ arises from a dimension two operator in the electric theory. Now the classical R-symmetry is
\be
R_{\Phi_{11}}=2\;\;,\;\;R_{q_1}=R_{\t q_1}=0\;\;,\;\;R_{\Phi_{00}}=1\;\;,\;\;R_{q_0}=R_{\t q_0}=\frac{1}{2}\;\;,\;\;R_{\Phi_{10}}=R_{\Phi_{01}}=\frac{3}{2}\,.
\ee


The stabilization of $\Phi_{00}$ is obtained from (\ref{eq:V2log}) and (\ref{eq:def2}), yielding
\be
|\Phi_{00}| = \sqrt{2 \t N_c (N_{f,1}-\t N_c)} \frac{h^2}{16 \pi^2} \frac{h \mu^2}{h^2 \muphi} \,\log \frac{|\Phi_{00}|^2}{\mu^2}
\ee
which can be solved iteratively for $|\Phi_{00}|$. As before, the R-symmetry is spontaneously broken and the phase of $\Phi_{00}$ is the massless R-axion. We will see in \S \ref{sec:pheno} that it acquires a mass after introducing an R-symmetry breaking constant to cancel the cosmological constant, without generating CP violating phases.  In this range we may integrate out the $q_0$ fields first, giving rise to a superpotential for $Z_0$ and $L$
\be
W \supset - h \frac{\mu^2}{\Phi_{00}} Z_0 \t Z_0 - h \frac{\mu}{\Phi_{00}} (Z_0 \t L \t \rho+ \rho L \t Z_0)-h \frac{1}{\Phi_{00}} \rho L \t L \t \rho\,.
\ee
The fields $Z_0$ have light masses of order $h \mu^2/\Phi_{00}$, while the $L$-pseudo-moduli have interactions with the $\rho$ fields that are suppressed by $h\mu/\Phi_{00}$. This explains the CW mass (\ref{eq:cwL2})~\cite{Giveon:2008ne}.

We see that the large-field approximation can be done self-consistently for
\be\label{eq:cond2}
h\muphi \lesssim \frac{h^2}{16 \pi^2}\mu\,.
\ee
This is natural given the dynamical origin of $\mu$ and $\muphi$ in the electric theory.  Finally, from the superpotential deformation (\ref{eq:def2}), we conclude that two-loop effects induce a nonzero F-term
\be\label{eq:FtermPhi}
\langle W_{\Phi_{00}}\rangle \approx h^2 \muphi \langle \Phi_{00}\rangle \sim \frac{h^2}{16 \pi^2} h \mu^2\,.
\ee

\vskip 2mm

We also need to ensure that, in both regimes, the metastable vacua are long-lived and robust against microscopic corrections. These points will be addressed in \S \ref{subsec:stability}, with the conclusion that small $\muphi/\mu$ and $\lambda$ guarantee stability of these vacua.

This ends our analysis of the quiver gauge theories that can be obtained at long distance from SQCD with massless and massive quarks. Such theories could have different applications. In particular, models of gauge mediation can be constructed by weakly gauging different flavor groups. The three-node quiver of Figure \ref{fig:quiver2} illustrates the rich set of possibilities.

\section{Gauge mediation with dynamical messengers}\label{sec:dyn-mess}

In the rest of the paper we apply the previous results to construct models of gauge mediation with `dynamical' messengers. By this we mean that the messenger sector and its interactions are not introduced in an ad-hoc manner, but rather are naturally generated by the same confining dynamics that breaks supersymmetry.

Starting from the quiver of Figure \ref{fig:quiver2}, we accomplish this by weakly gauging
\be\label{eq:embed}
G_{SM} \subset SU(N_{f,0})\,.
\ee
The hidden-sector fields charged under the SM are then $(q_0, \t q_0)$, $(Z_0, \t Z_0)$, $(L, \t L)$ and $\Phi_{00}$. Their SM quantum numbers are obtained by decomposing the (anti)fundamental and adjoint representations of $SU(N_{f,0})$ into SM representations according to the embedding (\ref{eq:embed}) that we choose. A simple example will be presented below.

Unfortunately, this model gives tachyonic two loop masses to the SSM sfermions, producing a phenomenological disaster. The source of this problem is that some of the hidden-sector fields charged under $G_{SM}$ have an unsuppressed positive supertrace $\rm{Str}\,M^2$. In models where the messengers have nonvanishing supertrace, the sfermion mass is, schematically,~\cite{Poppitz:1996xw}
\be
m_{\t f}^2 \sim \left(\frac{g_{SM}^2}{16 \pi^2} \right)^2\,   \left( \frac{F^2}{M^2}  -{\rm Str} M^2\, \log\frac{\Lambda_{{\rm cutoff}}}{M} \right) \,.
\ee
Here $M$ denotes the supersymmetric mass and $F$ gives the splittings in the bosons; for simplicity we have assumed small $F/M^2$. For the spectrum described in \S \ref{sec:different}, we find tachyonic contributions proportional to $m_{CW}^2$. They come from loops of $\rho$ and $Z$ that induce soft masses for $Z_0$ and $L$, which in turn are charged under $G_{SM}$ and would act as messengers. These effects are absent in minimal gauge mediation.

In fact, this is quite a generic problem for strongly-coupled models of direct mediation or single-sector models where $\rm{Str}\,M^2$ is not suppressed by factors of $F/M^2$~\cite{ArkaniHamed:1997jv,ArkaniHamed:1997fq}. In order to have a realistic model, we need to lift the hidden sector matter that has unsuppressed supertrace. Fortunately, here we can deal with such problems by introducing additional singlets $S$ and coupling them to the unwanted mesons (pairs of quarks) in the magnetic (electric) theory. This technique was used in the context of single-sector models to lift unwanted composite matter~\cite{Franco:2009wf,Craig:2009hf,Behbahani:2010wh}.\footnote{Other applications include, for instance the solution to $\mu/B_\mu$ in~\cite{SchaferNameki:2010iz}, and the recent proposal for direct gaugino mediation by~\cite{Green:2010ww}.}

\subsection{SQCD plus singlets}

From the previous discussion, we need to lift the components of $Z_0$, $L$ and $\Phi_{00}$ that are charged under the SM. Besides avoiding tachyonic sfermions, decoupling the additional matter will also allow us to achieve perturbative gauge coupling unification in $G_{SM}$. The overall singlet in $\Phi_{00}$ will not be lifted because it contributes F-term splittings to the messenger fields $q_0$. Similarly, we need to have at least one $L$-singlet in order to transmit supersymmetry breaking from the $SU(N_{f,1})$ node to the messenger node $SU(N_{f,0})$ at the two loop order.

For instance, we can choose $N_{f,0}=6$ massless quarks $(Q_a, \t Q_a)$, and embed $G_{SM} \subset SU(6)$ by
\be\label{eq:ex-embed}
Q_a \sim \mathbf 1 + \mathbf 5\;\;,\;\;\t Q_a \sim \mathbf 1 + \overline {\mathbf  5}\,.
\ee
Thus
\bea\label{eq:ex-embed2}
Z_0 & \sim & \t N_c(\mathbf 1 + \overline{\mathbf 5})\;\;,\;\;L \sim N_{f,1}(\mathbf 1 + \overline{\mathbf 5})\nonumber\\
q_0 & \sim & \t N_c(\mathbf 1 + \overline{\mathbf 5})\;\;,\;\;\Phi_{00} \sim \mathbf 1 + \mathbf 1 + \mathbf{24} + \mathbf 5 + \overline{\mathbf 5}\,,
\eea
and similarly for the conjugates.

We introduce singlets $S_{\mathbf R}$ in representations $\mathbf R$ of $G_{SM}$ and couple them to the electric quarks by
\be\label{eq:singletW}
W \supset y\,\sum_{\mathbf R}\,S_{\mathbf R}(Q \t Q)_{\overline{\mathbf R}}\,.
\ee
Here $(Q \t Q)_{\overline{\mathbf R}}$ runs over all the unwanted representations in (\ref{eq:ex-embed2}). Below the dynamical scale $\Lambda$ the cubic interaction becomes relevant, giving masses of order $y \Lambda$ to the unwanted matter. Notice that we have not introduced dimensionful parameters in this process. Adding the correct number of spectators and writing down the most general cubic couplings consistent with gauge invariance lifts all the fields in (\ref{eq:ex-embed2}) that have nonzero $G_{SM}$ quantum numbers. Furthermore, by assigning different discrete charges to the electric quarks
we can ensure that under a generic superpotential (\ref{eq:singletW}) the singlets in $L$ and $\t L$ and the overall singlet in $\Phi_{00}$ survive while the other ones are lifted.

Let us now consider the infrared magnetic description with, for concreteness, the embedding (\ref{eq:ex-embed2}). It is useful to show explicitly the matter content and nonabelian symmetries:
\begin{center}
\begin{tabular}{c|ccccc}
&$SU(\t N_c)_G$&$SU(\t N_c)$&$SU(N_{f,1}- \t N_c)$&$SU(5)_{SM}$  \\
\hline
&&&&\\[-12pt]
$\chi$&$\Box$& $\overline \Box$&$1$&$1$  \\
$\t \chi$&$\overline \Box$& $\Box$&$1$&$1$  \\
$\rho$&$\Box$&$1$&$ \overline \Box$&$1$  \\
$\t \rho$&$\overline \Box$&$1$&$ \overline \Box$&$1$  \\
$Z$&$1$&$\Box$&$ \overline \Box$&$1$  \\
$\t Z$&$1$&$\overline \Box$&$ \overline \Box$&$1$  \\
$Y$&$1$& $\textrm{adj}+1$&$1$&$1$  \\
$X$&$1$&$1$&$\textrm{adj}+1$&$1$  \\
$q_0$&$\Box$&$1$&$1$&$\overline \Box + 1$  \\
$\t q_0$&$\overline \Box$&$1$&$1$&$\Box+1$  \\
$L$&$1$&$1$&$\Box$&$1$  \\
$\t L$&$1$&$1$&$\overline\Box$&$1$  \\
$\Phi_{00}$&$1$&$1$&$1$&$1$
\end{tabular}
\end{center}
The superpotential of the magnetic theory is given by (\ref{eq:Wgen}). Decoupling the heavy states lifted by the singlets yields
\bea\label{eq:Wnew}
W_{mag}&=& - h \mu^2 Y + h \chi Y \t \chi - h \mu^2 X + h \rho X \t \rho + h (\rho \t Z \t \chi + \chi Z \t \rho)+\nonumber\\
&+& h \rho L \t q_0 + h q_0 \t L \t \rho + h q_0 \Phi_{00} \t q_0 + f(\Phi_{00})
\eea
where we have omitted color and flavor indices. The function $f(\Phi_{00})$ denotes the quadratic or cubic deformations discussed in \S \ref{subsec:2-loop},
\be
f(\Phi_{00}) \equiv \frac{1}{2}h^2\muphi\, \Phi_{00}^2\;,\;\textrm{or}\;\;f(\Phi_{00}) \equiv \frac{1}{3}\lambda \Phi_{00}^3\,.
\ee

Lifting mesons can have important consequences on the vacuum structure, particularly if some of these fields were playing an important role in the breaking of supersymmetry by the rank condition.\footnote{A version of this issue was encountered for instance in single sector models~\cite{Franco:2009wf,Craig:2009hf}; the change in vacuum structure and appearance of new metastable states was discussed in detail by~\cite{Behbahani:2010wh,SchaferNameki:2010iz}.} Therefore, we next analyze the vacuum structure of the new superpotential (\ref{eq:Wnew}).

\subsection{Supersymmetry breaking in the deformed theory}\label{subsec:dsb-SSQCD}

First, the supersymmetry breaking dynamics from the $N_{f,1}$ node of the quiver theory (see Figure \ref{fig:quiver1}) is not modified. This is because the extra singlets only couple to fields in the $N_{f,0}$ node or to the link fields $L$ connecting both nodes. In more detail, the tree level F-terms for $q_1$, $\t q_1$ and $\Phi_{11}$ (namely the fields appearing in the first line of (\ref{eq:Wnew})) imply that supersymmetry is broken by the rank condition,
\be
\langle \chi \t \chi \rangle = \mu^2 \mathbf 1_{\t N_c \times \t N_c}\;\;,\;\;W_X = - h \mu^2 \,\mathbf 1_{(N_{f,1}- \t N_c) \times (N_{f,1}- \t N_c)}\,,
\ee
as in (\ref{eq:ISSvac}). $(Y, \rho, \t \rho, Z, \t Z)$ are stabilized at the origin (except for the NGBs discussed in \S \ref{subsec:dsb}); $X$ and $\rm{Re}\,\tr(\chi - \t \chi)$ are pseudomoduli. 

The fields $(L, \t L)$ are flat at tree level; as before, they receive a one loop potential (from integrating out $\rho$ and $Z$) that stabilizes them at the origin. Next, $\Phi_{00}$ is flat at one loop. Even though the bifundamentals $(Z_0, \t Z_0)$ are no longer part of the low energy theory, one can show that $\Phi_{00}$ gets a nonzero two loop potential, similar to the one discussed in \S\ref{subsec:2-loop}, from diagrams containing $(L, \t L)$. This is the reason why such singlets were not lifted. The  $(L, \t L)$ singlets give $\Phi_{00}$ a tachyonic mass near the origin and a logarithmic dependence for $\Phi_{00}\gtrsim h \mu$.

The important difference between the model with singlets and that of \S \ref{sec:different} is that, having lifted $(Z_0, \t Z_0)$, the magnetic quarks $(q_0, \t q_0)$ are no longer stabilized at the origin by the expectation value $\chi \t \chi = \mu^2$. In fact, these fields are massless if $\Phi_{00}=0$, so they need to be included when looking for supersymmetry breaking vacua. The relevant potential is
\be\label{eq:VqPhi}
V= |h q_0 \t q_0+ f'(\Phi_{00})|^2+ h^2 |\Phi_{00}|^2(|q_0|^2+|\t q_0|^2)+ V_{CW}^{(2)}(|\Phi_{00}|)+V_D\,.
\ee
Here $V_D$ denotes the D-term potential from the magnetic gauge group and weakly gauged flavor groups; minimizing these contributions requires $|q_0|=|\t q_0|$.

In order to understand the vacuum structure of (\ref{eq:VqPhi}), let us first set the small deformation $f(\Phi_{00})=0$. It is not hard to see that critical points can only exist for $q_0 = \t q_0=0$ and hence the runaway for $\Phi_{00}$ is not modified. In particular, once the two loop instability forces $\Phi_{00}$ to condense, the magnetic quarks will get a positive mass squared. Turning on a small enough deformation $f(\Phi_{00})$ (the case of interest for us), does not modify this conclusion. Quantitatively, the smallness of $f$ requires $\muphi \lesssim \frac{\mu}{16 \pi^2}$ for $f(\Phi) = \frac{h^2}{2} \muphi \Phi_{00}^2$, or $\lambda \lesssim h$ for the cubic case $f(\Phi_{00}) = \frac{\lambda}{3} \Phi_{00}^3$. 

To summarize, we find that the theory with singlets and couplings (\ref{eq:singletW}) has a metastable vacuum where the $N_{f,1}$ node breaks supersymmetry spontaneously.  SUSY breaking is then transmitted to the $N_{f,0}$ node, where $\Phi_{00}$ is stabilized by a competition between two loop effects and the superpotential deformation $f(\Phi_{00})$. Up to numerical coefficients, the vacuum is the same as in \S \ref{subsec:2-loop}. No new vacua arise for a small enough deformation.

\subsection{Stability of the vacuum}\label{subsec:stability}

The last step in demonstrating the consistency of the models is to show that the vacua are (meta)stable, both against decay to other vacua and against microscopic corrections from the electric UV completion. Let us discuss the lifetime first.

As there are no other metastable vacua, the lifetime is constrained by decay to the supersymmetric vacua.  The lifetime calculation is roughly the same as done in~\cite{Intriligator:2006dd}.  There is a dynamically generated superpotential
\be
W_{dyn} = \t N_c  \left(h^{N_f} \frac{\det \Phi}{\Lambda^{N_f-3 \t N_c}}\right)^{1/\t N_c}
\ee
The most efficient path is to go to $q=\t q=0$ then proceed along the mesonic directions.  The potential is schematically,
\be
V \sim N_{f,1} \left|\Lambda^{3-N_f/\t N_c}\Phi_{00}^{N_{f,0}/\t N_c} \Phi_{11}^{N_{f,1}/\t N_c-1 } - \mu^2\right|^2 + N_{f,0} \left|\Lambda^{3-N_f/\t N_c}\Phi_{00}^{N_{f,0}/\t N_c-1} \Phi_{11}^{N_{f,1}/\t N_c } - f'(\Phi_{00}) \right|^2\,.
\ee
The major contribution to the potential comes from the first term.  To travel to a point where $V = V_{non SUSY} = (N_{f,1}-\t N_c) \mu^4$ requires setting 
\be
\label{eq:curve}
\Phi_{00}^{N_{f,0}/\t N_c} \Phi_{11}^{N_{f,1}/\t N_c-1 } \sim \mu^2 \Lambda^{N_f/\t N_c -3}
\ee

The bounce action is given by a triangle approximation and is roughly $\frac{\Delta \Phi^4}{V}$,  where $\Delta \Phi$ is the minimum distance from our metastable to the curve defined in (\ref{eq:curve}).  In order to have a parametrically long-lived vacuum this distance should be much greater than $\mu$. This places the bounds 
\be
\frac{\muphi}{\mu} \gg \frac{1}{16 \pi^2} \left(\frac{\mu}{\Lambda}\right)^{\frac{N_f - 3 \t N_c}{N_{f,0}}}\;,\;\lambda \gg \frac{1}{16 \pi^2} \left(\frac{\mu}{\Lambda}\right)^{\frac{N_f - 3 \t N_c}{N_{f,0}}} \,.
\ee
These constraints will be satisfied in the concrete examples of \S \ref{sec:pheno}.

Finally, we address the stability against microscopic corrections. Integrating out short distance modes at the scale $\Lambda$ produces K\"ahler potential corrections
\be\label{eq:delK}
\delta K = c \frac{|\Phi|^4}{|\Lambda|^2} + \ldots
\ee
in terms of an incalculable constant $c$. As long as $\mu/\Lambda \ll 1$, microscopic corrections to the stabilization of $X$ are negligible, but the constraints from $\Phi_{00}$ are stronger because its potential depends on small two loop effects. Requiring the $X X ^\dag/\Lambda^2$ corrections to be smaller than the light $\Phi_{00}$ mass sets, for the model with quadratic deformation,
\be\label{eq:upper1}
\frac{\mu^2}{\Lambda}\ll h \muphi\,.
\ee
Recall that in the quadratic case there is a lower bound on $\mu$ given by (\ref{eq:cond2}). On the other hand, for the case of a cubic deformation, the condition is
\be
\mu \ll \frac{h^2}{16 \pi^2} \Lambda\,.
\ee

\subsection{Microscopic lessons for messenger gauge mediation}

In this way, we obtain a consistent model of supersymmetry breaking with a messenger sector $(q_0, \t q_0)$ that couples to the supersymmetry breaking dynamics via
\be\label{eq:Wmess}
W_{mess}= h  q_0  \Phi_{00} \t q_0\,.
\ee
This gives rise to supersymmetric masses and F-term splittings
\be\label{eq:messM}
M = h \langle \Phi_{00} \rangle\;\;,\;\;-F^* = h \langle f'(\Phi_{00}) \rangle\,.
\ee
Decoupling the messengers does not appreciably affect the supersymmetry breaking structure of the theory.

The fields $(q_0, \t q_0)$ also couple to the supersymmetry breaking node $N_{f,1}$ via the gauge bosons of the magnetic gauge group. As a result, they will acquire soft masses at two loops. These effects are similar to the gauge-mediated masses for the MSSM sfermions, with the difference being that the gauge group is supersymmetrically higgsed at the scale of supersymmetry breaking. A related generalization of gauge mediation was studied by~\cite{Gorbatov:2008qa}. However, since the magnetic gauge coupling $\t g$ is IR free, these effects are typically smaller than the masses (\ref{eq:messM}). By requiring $\t g < h$, these soft masses are negligible.

Having constructed a UV completion of messenger gauge mediation where the supersymmetry breaking and messenger fields are unified into a single sector, it is useful to adopt a more general perspective and point out some basic lessons. If $\mc O_m$ denotes an operator
made of the messengers $(q_0, \t q_0)$, and $\mc O_h$ is a function of the remaining hidden sector fields, the interactions between both sectors are of the form
\be\label{eq:generalGM}
L \supset \int d^2 \theta\,\mc O_h \mc O_m + \int d^4 \theta\,\mc O_h' \mc O_m' + \textrm{c.c.}\,.
\ee
This is a microscopic realization of ``general messenger gauge mediation''~\cite{Dumitrescu:2010ha}. Our setup is dominated by the superpotential coupling (\ref{eq:Wmess}), and K\"ahler potential interactions from magnetic gauge bosons are subdominant. Seiberg duality gives a weakly coupled description, and the messengers decouple from the supersymmetry breaking node when $\t g \to 0$ and $h \to 0$.

One lesson from our microscopic construction is that the unification of the supersymmetry breaking and messenger sectors into a strongly coupled single sector eliminates many of the distinctions between direct and indirect mediation. Indeed, we can interpret the electric theory as a model of direct mediation, while the IR limit gives indirect mediation. Both mechanisms appear then related by duality.

We also found that messengers can generically obtain unsuppressed soft masses, and presented an example of this in \S \ref{sec:different}. A large positive supertrace then leads to tachyonic MSSM sfermions. In order to detect and avoid these tachyons it was crucial for us to have a weakly coupled dual description. The dangerous contributions were decoupled by adding spectator fields.

Recall that in bottom-up approaches to gauge mediation, one assumes a hidden sector with a spurion field that has fixed expectation value and F-term, and then couples it to messengers as in (\ref{eq:Wmess}). An important question is whether such a coupling can change the supersymmetry breaking vacuum. Any UV completion of messenger gauge mediation needs to face this question. The discussion in \S\ref{subsec:dsb-SSQCD} was related to this point.

The `visible' F-term, controlled by $W_{\Phi_{00}}$, is much smaller than the `hidden' F-term $W_X$ that triggers supersymmetry breaking in the $N_{f,1}$ node. Supersymmetry breaking is transmitted to $\Phi_{00}$ only at two loops.\footnote{The situation is reminiscent of the cascade model presented in~\cite{Ibe:2010jb} which, unfortunately, is not fully calculable. See also~\cite{Evans:2011pz}.}  This realizes dynamically some of the ideas in the original works~\cite{Dine:1993yw,Dine:1994vc,Dine:1995ag} and leads to interesting consequences for the phenomenology of the models. While the original direct mediation construction of ISS was restricted to low scale supersymmetry breaking, this is not necessarily the case in our messenger gauge mediation constructions and leads to the intriguing possibility of a heavy GeV-range gravitino. Specific parameter choices with a realistic phenomenology will be presented in \S \ref{sec:pheno}.

\section{Phenomenology of the models}\label{sec:pheno}

In this section we discuss the basic phenomenological features of our models. Before proceeding, it is necessary to point out that we have not addressed $\mu/B_\mu$. It would be interesting to understand whether some of the fields in the quiver theory can be used for this purpose, perhaps along the lines of~\cite{SchaferNameki:2010mg}. Also, we present the soft masses at the messenger scale, while a more detailed analysis of the signals requires running down to the TeV scale.

One important consequence of the introduction of spectator fields is that it is possible to achieve perturbative gauge coupling unification. Below the dynamical scale $\Lambda$, the messenger index is given by $\t N_c$, the rank of the magnetic gauge group. However, at an energy scale of order $\Lambda$, the composite fields that couple to the spectators start contributing to the $G_{SM}$ running and rapidly give rise to a Landau pole. Therefore, to maintain perturbative unification we need to choose
\be
\Lambda \gtrsim M_{GUT}\;,\;\; \t N_c \lesssim \frac{150}{\log(M_{GUT}/M)}
\ee
(see~\cite{Giudice:1998bp}).

It is also important to point out that the class of models presented here have comparable gaugino and sfermion masses. The conclusions of~\cite{Komargodski:2009jf} on small gaugino masses are evaded because supersymmetry is broken by radiative corrections and not by a tree-level superpotential term. Furthermore, the field $\Phi_{00}$ whose F-term provides the MSSM soft parameters is not a pseudo-modulus.

\subsection{Model with $\Phi_{00} \ll \mu$}

This class of models has messengers with
\be
\frac{|F|}{|M|} \approx \frac{h^2}{16 \pi^2} \sqrt{\t N_c (N_{f,1}-\t N_c)} h \mu .
\ee
The cubic coupling $\lambda$ cancels out from this ratio. 

Gauginos and sfermions naturally obtain masses of the same order. In particular, from the one loop gaugino mass
\be
m_{\lambda_a} \approx \frac{g_a^2}{16 \pi^2} \t N_c \frac{F}{M}\,,
\ee
requiring a gluino of mass $\sim 1$ TeV sets
\be
F_X^{1/2}= \sqrt{h}\mu \approx h^{-5/2}\times10^7\,\rm{GeV}\,.
\ee
Typically $h \sim \mc O(1)$ and these are models of low scale supersymmetry breaking.

The gravitino mass is determined by the dominant F-term,
\be
m_{3/2} \approx \frac{F_X}{\sqrt{3} M_{pl}} \approx h^{-5}\times100\;\rm{keV}\,,
\ee
where $M_{pl} \approx 2.4 \times 10^{18}\,\rm{GeV}$ is the reduced Planck mass. A gravitino with a mass larger than a few keV can account for all the dark matter but requires a non-standard cosmology to avoid overclosing the universe~\cite{Viel:2005qj,Feng:2010ij}.

The leading contribution to the R-axion mass comes from anomaly-mediated A-terms~\cite{Randall:1998uk},
\be
V = \frac{\gamma_{\Phi_{00}}}{2} m_{3/2} \lambda \Phi_{00}^3 + \text{h.c.}
\ee
where the anomalous dimension of $\Phi_{00}$ is $\gamma_{\Phi_{00}}\sim h^2/16\pi^2$
This gives an axion mass 
\be
m_a^2 \sim \frac{h^2}{16 \pi^2} m_{3/2} \lambda \Phi_{00} \sim h^{-3} \times 10^{-1} \text{GeV}^2\,.
\ee
Then
\be
\frac{F}{M}=- \lambda \bar{\Phi}_{00}^3
\ee
is real and no phases for the gaugino masses are generated.

For $h \lesssim 1$ the phenomenology of this model is that of minimal gauge mediation with a heavy gravitino.  The effect of a heavy gravitino is to make the NLSP stable on collider scales.  For  $\t N_c \gtrsim 3$, the stau is the NLSP (if the higgsino is not lighter than the gauginos), while for smaller values of $\t N_c$, the neutralino is the NLSP. If the NLSP is a neutralino, then the collider signatures are that of typical supersymmetry with a stable neutralino.  Having a stau NLSP or co-NLSP yields an exotic signature where a charged massive particle makes it all the way through the detector. 

An interesting feature is that increasing the coupling $h$ (related to the ratio of electric and magnetic scales of SQCD) decreases the scale of supersymmetry breaking. Going to a regime where the decay of the NLSP is prompt ($F_X^{1/2} \lesssim 10^5\,\rm{GeV}$) may require fine-tuning and/or loss of perturbativity. However, a moderate increase in $h$ can reach an intermediate regime where NLSP decays produce displaced vertices. This is very interesting experimentally. It would be worth to study this range in more detail.

In either case, the hidden sector particles are always in the multi-TeV range and are therefore unobservable at the LHC.

\subsection{Model with $\Phi_{00} \gg \mu$}\label{subsec:largePhi}

The model based on the deformation $W \supset \frac{h^2}{2} \muphi \Phi_{00}^2$ exhibits a rich phenomenology, partly because $\muphi$ determines the soft parameters. In this case the messengers have
\be
\frac{F}{M}= h^2 \muphi\,,
\ee
and $F \ll M^2$; see (\ref{eq:FtermPhi}). Having a gluino around 1 TeV sets
\be
h^2 \muphi \sim\,10^5\;\rm{GeV}\,.
\ee

The hidden sector F-term $F_X^{1/2} = \sqrt{h} \mu$ is not fixed by the visible masses; nevertheless $\mu$ needs to satisfy certain constraints for the consistency of the model. A lower bound on $F_X$ comes from (\ref{eq:cond2}). We also found an upper bound (\ref{eq:upper1}) by requiring that microscopic corrections to the K\"ahler potential are parametrically small. A stronger upper bound is obtained by requiring that Planck suppressed operators
\be
\Delta L  =\int d^4 \theta\, \Phi_{SM}^\dag \Phi_{SM}\,\frac{X^\dag X}{M_{pl}^2}
\ee
do not produce dangerous FCNCs. These operators can induce flavor violating sfermion masses that need to be suppressed by~\cite{Gabbiani:1996hi}
\be
\frac{\Delta m^2_{\t f}}{m_{\t f}^2} \lesssim 10^{-3}\,.
\ee
In total, the constraint on the hidden sector $F_X= h^2 \mu$ reads
\be\label{eq:Fxwindow}
h^{-5/2} \times 10^7\,{\rm GeV} < F_X^{1/2} < 3 \times 10^9\,\rm{GeV}
\ee
assuming a gluino around 1 TeV and the lightest sfermion around 100 GeV.

The gravitino mass is predicted to be in the window
\be\label{eq:gr-range}
h^{-5}\times100\;{\rm keV}\, < m_{3/2}< 1\,\rm{GeV}\,.
\ee
The gravitino can account for all cold dark matter but again a non-standard cosmology is required. The R axion receives a mass from an explicit R symmetry breaking constant introduced to cancel the cosmological constant. Following~\cite{Bagger:1994hh} obtains
\be
m_a^2 \approx h^2 \muphi m_{3/2}  \gtrsim h^{-5}\times 10 \, {\rm GeV}^2
\ee
which avoids astrophysical bounds for a perturbative $h$. As before, no phases for the gaugino masses are generated.

The phenomenology near the lower end of $F_X$ is similar to the $\Phi_{00} \ll \mu$ model.  The main difference is that the gravitino mass is not fixed by the MSSM soft masses. In the range (\ref{eq:gr-range}) the heavy gravitino implies that the NLSP escapes the detector and, again, a slepton NLSP could have interesting consequences.

This leaves open the intriguing possibility of modifying the model to increase $\mu$ beyond (\ref{eq:Fxwindow}), without altering the TeV spectrum of soft masses. In this case, the gravitino would not be the LSP! We will comment more on this possibility in the conclusions.

\section{Conclusions}\label{sec:concl}

SQCD in the presence of multiple mass terms has a rich set of dynamics.  In this work, we showed that this theory provides a microscopic realization of gauge mediation with messengers. The supersymmetry breaking and messenger sectors are unified into a single sector, and the ad hoc appearance of messengers and their interactions is eliminated. We presented fully consistent models of gauge mediation including both the supersymmetry breaking sector and messengers.

Our constructions are calculable, have metastable supersymmetry breaking, break the R-symmetry spontaneously and lead to realistic gaugino and sfermion masses. Other phenomenological features include a heavy gravitino and the possibility of displaced vertices from the NLSP.

In the simplest realization of this approach there are unsuppressed supertraces that cause tachyonic sfermions. This could be an ubiquitous problem of dynamical models of gauge mediation. Having a weakly coupled dual gave us the opportunity of explicitly addressing this; our solution involved adding spectator fields to decouple the undesired particles. Spectators were also important for achieving perturbative gauge coupling unification. The use of spectators adds some degree of arbitrariness, and it would be nice to find models where they are not needed.

An interesting case appeared in the model of \S \ref{subsec:largePhi}. The parameter which governed the MSSM spectrum was independent of the dominant SUSY breaking and the gravitino mass is a completely independent parameter.  The upper bound on the SUSY breaking F term is provided only by Planck suppressed operators causing FCNCs. If conformal sequestering is used to suppress these effects, then there is a window where the gravitino is not the LSP and can decay to MSSM particles! This would lead to the intriguing possibility of gauge mediation with high scale of supersymmetry breaking, and a neutralino dark matter (see e.g.~\cite{Dudas:2008eq,Craig:2008vs}). Conformal sequestering could be implemented already in our single sector theory, by adding an extra adjoint field~\cite{Schmaltz:2006qs}.

\section*{Acknowledgments}
We are grateful to R.~Essig and S.~Kachru for very helpful discussions on gauge mediation and phenomenology. We would also like to thank
N.~Craig,
R.~Essig,
S.~Kachru,
S.~Franco,
A.~Nacif,
S.~Schafer-Nameki,
C.~Tamarit  and J.~Wacker
for helpful comments on our work. We are supported by the US DOE under contract number DE-AC02-76SF00515 at SLAC.


\bibliographystyle{JHEP}
\renewcommand{\refname}{Bibliography}
\addcontentsline{toc}{section}{Bibliography}
\providecommand{\href}[2]{#2}\begingroup\raggedright
\end{document}